\documentclass[oneside]{article}
\usepackage[T1]{fontenc}
\usepackage{authblk}
\usepackage{float}
\usepackage{amsmath}
\usepackage{graphicx}
\usepackage{xfrac}
\usepackage[english]{babel}
\usepackage{lmodern}
\usepackage[T1]{fontenc}
\usepackage[latin9]{inputenc}

\usepackage{csquotes}

\usepackage{mathtools}
\usepackage{graphicx}
\usepackage{esint}
\usepackage[unicode=true,
 bookmarks=false,
breaklinks=false,pdfborder={0 0 1},backref=false,colorlinks=false]
 {hyperref}
\usepackage[a4paper, total={210mm,297mm},margin=2.0cm]{geometry}

\usepackage{dcolumn}
\usepackage{bm}
\usepackage{lineno}
\usepackage{xcolor}
\usepackage{algpseudocode} 
\usepackage[]{algorithmicx}
\usepackage{varwidth}

\title{Mesoscale modeling of near-contact interactions for complex flowing interfaces}

\author[1]{Andrea Montessori \thanks{Electronic address: \texttt{a.montessori@iac.cnr.it}; Corresponding author}}
\author[1]{Marco Lauricella}
\author[3]{Nicola Tirelli}
\author[4,1,5]{Sauro Succi}

\affil[1]{Istituto per le Applicazioni del Calcolo CNR, via dei Taurini 19, Rome, Italy}
\affil[2]{North West Centre for Advanced Drug Delivery (NoWCADD), Division of Pharmacy and Optometry, School of Health Sciences, Medicine and Health, Stopford Building, Manchester M13 9PT, United Kingdom}
\affil[3]{Laboratory of Polymers and Biomaterials, Fondazione Istituto Italiano di Tecnologia, 16163 Genova, Italy}
\affil[4]{Center for Life Nano Science@La Sapienza, Istituto Italiano di Tecnologia, 00161 Roma, Italy}
\affil[5]{Institute for Applied Computational Science, Harvard John A. Paulson School of Engineering And Applied Sciences, Cambridge, MA 02138, United States}

\date{\today}

\begin{document}

\maketitle

\begin{abstract}
We present a mesoscale kinetic model for multicomponent flows, augmented with a short range forcing term, aimed at describing the combined effect of surface tension and near-contact interactions operating at the fluid interface level.
Such mesoscale approach is shown to i) accurately capture the complex dynamics of bouncing colliding droplets for different values of the main governing parameters, ii) predict quantitatively the effective viscosity of dense emulsions in micro-channels and iii) simulate the formation of the so-called \textit{soft flowing crystals} in microfluidic focusers. 
\end{abstract}

\section{Introduction}

A thorough knowledge of the dynamic interactions between fluid interfaces is paramount 
to a deeper understanding of a variety of natural processes and engineering applications,
such as combustion, microfluidic coating, food processing and many others.\\
The coalescence and/or repulsion between droplets or bubbles can be traced to the 
hydrodynamic drag originating from the relative motion of two fluid interfaces in near contact (\cite{davis1989lubrication,barnocky1989lubrication,rubin2017elastic,shi1994cascade,mani2010events}), and to the combined action of nanoscale attractive and/or repulsive forces, such as Van der Waals and electrostatic forces, steric interactions, hydration repulsion and depletion attraction (\cite{stubenrauch2003disjoining,bergeron1999forces}).

A wide body of theoretical and experimental work has elucidated the complex nature of the near contact interactions which develop within intervening liquid films: from the pioneering works of Gibbs and Marangoni on the thermodynamics of liquid thin films (see (\cite{bergeron1999forces}) for a comprehensive review) to the separate work of Derjaguin and Overbeek (\cite{derjaguin1940repulsive,verwey1999theory}) which culminated in the joint DLVO theory (after
its founders, Derjaguin, Landau, Verwey and Overbeek).\\
These seminal works laid down the foundations for describing a broad variety of complex flowing
systems, such as colloids, foams and emulsions, as well as flowing collections of droplets and bubbles
characterized by highly ordered and uniform, crystal-like structures, now known as \textit{soft flowing crystals} (\cite{marmottant2009microfluidics,garstecki2006flowing}).\\
From a numerical standpoint, the direct introduction of near interaction forces at a molecular level reflects the need of solving simultaneously six spatial decades: from millimeters, i.e. the typical size of microfluidic devices, down to nanometers (and below), namely the relevant spatial scale of contact forces. This is far beyond the capability of any foreseeable computer (\cite{montessori2018mesoscale}), hence placing a high premium on coarse-grained, mesoscale representations of near-contact forces, 
capable of retaining computational viability without compromising the essential physics.

Of course, the success of such mesoscale strategy hinges crucially on the universality of the
underlying physics, i.e. its dependence on suitable dimensionless parameters measuring the
relative strength of the interactions, rather than on the details and strength of the interactions themselves (\cite{succi2015lattice}).
Failing such universality, a genuine microscopic approach cannot be helped, thus hampering the possibility to reach up to the scales of the full device.

In the LB framework, immiscible fluids were first modelled by \cite{gunstensen1991lattice}.These Authors developed a lattice Boltzmann model, augmented with a forcing term in accordance with the local color gradient, giving rise to the interface tension  and a segregation step. In \cite{latva2005diffusion}, the Authors corrected the segregation rule proposed by Gunstensen, to avoid the pinning problem that affected the Gunstensen model, allowing the fluids to moderately mix and to keep the color distribution symmetric with respect to the color gradient. The reason for lattice pinning is that at the sites where it happens, all of the particles of one kind are sent to the same direction and hence they cannot move from one site to another. 
More recently  the model was further improved to simulate high density and viscosity ratio (\cite{leclaire2012numerical, leclaire2017generalized,saito2017lattice}).
\\
\textcolor{black}{In addition to this, there have been several attempts to model interface interactions in amphiphilic fluids within the LB framework (see for example \cite{nekovee2000lattice,love2001three,Chen20002043}). These  models aim at directly simulating the effect of a surfactant by evolving two sets of distribution functions, allowing to  take into account the presence of an amphiphilic fluid at the interface between two liquid phases.
For a comprehensive review of multiphase and multicomponent LB models (\cite{swift1995lattice, philippi2012lattice, guo2005finite, shan1993lattice, hedoolen1998}) , the interested reader is referred to \cite{huang2015multiphase}}.

In this paper, we present a lattice Boltzmann (LB) based approach for multicomponent flows, based on the color-gradient model \cite{leclaire2012numerical},  augmented with an additional forcing term which is aimed at representing the effects of the near-contact forces operating at the fluid interface level.
The proposed model, based   is shown to accurately capture the collision outcomes between bouncing droplets for different values of the governing parameters, to predict the effective viscosity of dense emulsions in channels and to effectively simulate the evolution of \textit{soft flowing crystals} in flow focuser devices.\\
The paper is organized as follows. In Section \ref{method} the lattice Boltzmann equation with the BGK collisional operator is described, together with the color gradient model and the regularization algorithm for simulating multicomponent fluids. The augmented LB model for repulsive near contact interactions is discussed in details in subsection \ref{NCI}. Section \ref{results} collects the main results of the paper. Finally, a summary is reported in section \ref{summary}.

\section{Method}\label{method}

Lattice Boltzmann models for non-ideal fluids come mainly in two families. 
The first one is based on heuristic assumptions
(\cite{leclaire2012numerical,leclaire2017generalized,korner2005lattice,montessori2018regularized,becker2009combined})
 while the second one,
 builds on the projection of the kinetic equation on a discrete set of microscopic velocities
 (\cite{swift1995lattice, philippi2012lattice,guo2005finite,shan1993lattice, montessori2017entropic, hedoolen1998}).
 For a more exhaustive review, see (\cite{huang2015multiphase,succi2018lattice}). 
In the following we provide
a brief introduction to the one which we found most suitable for the description of  \textit{flowing
crystals}, namely the regularized color gradient method (\cite{montessori2018regularized}).

\subsection{Regularized color gradient lattice Boltzmann model}
In the color gradient LB for multicomponent flows, two
sets of distribution functions are needed to track the evolution of the two fluid components, which occurs via a streaming-collision algorithm (for a comprehensive review on the lattice Boltzmann method, please refer to  (\cite{succi2018lattice,kruger2017lattice})):

\begin{equation} \label{CGLBE}
f_{i}^{k} \left(\vec{x}+\vec{c}_{i}\Delta t,\,t+\Delta t\right) =f_{i}^{k}\left(\vec{x},\,t\right)+\Omega_{i}^{k}( f_{i}^{k}\left(\vec{x},\,t\right)),
\end{equation}

where $f_{i}^{k}$ is the discrete distribution function, representing
the probability of finding a particle of the $k^{th}$ component at position $\vec{x}$ and time
$t$ with discrete velocity $\vec{c}_{i}$ . 

where and $i$ is the index running over the lattice discrete directions $i = 0,...,b$, where $b=26$ for a three dimensional 27 speed lattice (D3Q27).
\textcolor{black}{The lattice time step $\Delta t$ has been taken as $1$ (in lattice units) for convenience, which is a common practice in LB literature (see \cite{succi2018lattice}).}
\textcolor{black}{The density $\rho^{k}$ of the $k^{th}$ component is given  by the zeroth moment of the distribution functions:
\begin{equation}
\rho^{k}\left(\vec{x},\,t\right) = \sum_i f_{i}^{k}\left(\vec{x},\,t\right),
\end{equation}
The total fluid density is given by $\rho=\sum_k \rho^k$, while the total momentum of the mixture is
defined as the sum of the linear momentum of the two components:
\begin{equation}
\rho \vec{u} = \sum_k  \sum_i f_{i}^{k}\left(\vec{x},\,t\right) \vec{c}_{i}.
\end{equation}}
The collision operator can be split into three parts (\cite{gunstensen1991lattice,leclaire2012numerical,leclaire2017generalized}): 

\begin{equation}
\Omega_{i}^{k} = \left(\Omega_{i}^{k}\right)^{(3)}\left[\left(\Omega_{i}^{k}\right)^{(1)}+\left(\Omega_{i}^{k}\right)^{(2)}\right].
\end{equation}

In the above, $\left(\Omega_{i}^{k}\right)^{(1)}$ stands for the standard collisional relaxation (\cite{succi2001lattice}) \textcolor{black}{which reads $\left(\Omega_{i}^{k}\right)^{(1)}=\omega_{eff}(f_i^{k,eq} - f_i^k)$, where $\omega_{eff}=2/(6\bar{\nu} -1)$ is the effective relaxation parameter being $\bar{\nu}$ the viscosity at the interface between the two fluids which is computed as $\frac{1}{\bar{\nu}}=\frac{\rho_1}{(\rho_1+\rho_2)}\frac{1}{\nu_1} + \frac{\rho_2}{(\rho_1+\rho_2)}\frac{1}{\nu_2}$ ($\nu_1$ and $\nu_2$ are the kinematic viscosities of the two fluids in the bulk). 
The equilibrium distribution function of the $k^{th}$ component $f_i^{k,eq}$ is given by a low-Mach, second-order, expansion
of a local Maxwellian, namely\\
$f_i^{k,eq}=w_i \rho^k (1 + \frac{ \vec{c_i} \cdot \vec{u}}{c_s^2} +\frac{(\vec{c_i} \cdot \vec{u})^2}{2c_s^4} - \frac{\vec{u} \cdot \vec{u}}{2 c_s^2})$}.\\
$\left(\Omega_{i}^{k}\right)^{(2)}$ is the perturbation step (\cite{gunstensen1991lattice}), which 
contributes to the build up of an interfacial tension. Finally, $\left(\Omega_{i}^{k}\right)^{(3)}$ is the recoloring step (\cite{gunstensen1991lattice,latva2005diffusion}), which promotes the segregation  between species, so as to minimize their mutual diffusion.

In order to reproduce the correct form of the stress tensor (\cite{landau1959course}), the perturbation operator can be constructed by exploiting the concept of the continuum surface force (\cite{brackbill1992continuum}).
Firstly, the perturbation operator must satisfy the following  conservation constraints:

\begin{eqnarray} \label{consconstr}
\sum_i \left(\Omega_{i}^{k}\right)^{(2)}=0 \\
\sum_k \sum_i \left(\Omega_{i}^{k}\right)^{(2)} \vec{c}_i=0
\end{eqnarray}

By performing a Chapman-Enskog expansion, it can be shown that the hydrodynamic limit of Eq.\ref{CGLBE} is represented by a
set of equations for the conservation of mass and linear momentum:

\begin{eqnarray} \label{NSE}
\frac{\partial \rho}{\partial t} + \nabla \cdot {\rho \vec{u}}=0 \\
\frac{\partial \rho \vec{u}}{\partial t} + \nabla \cdot {\rho \vec{u}\vec{u}}=-\nabla p + \nabla \cdot [\rho \nu (\nabla \vec{u} + \nabla \vec{u}^T)] + \nabla \cdot \bm{\Sigma}
\end{eqnarray}

where $p=\sum_k p_k$ is the pressure and $\nu=c_s^2(\tau-1/2)$ is the kinematic viscosity of the mixture, being $\tau$ the single relaxation time and $c_s=1/\sqrt{3}$ the sound speed of the model (\cite{succi2001lattice,kruger2017lattice}).

\textcolor{black}{Note that the divergence of the stress tensor (last term in equation eq. 2.8), which is responsible for the build up of surface tension, acts only at the interface between the fluids (see eq. \ref{force})}

The stress tensor in the momentum equation is given by:

\begin{equation}
\bm{\Sigma}=-\tau\sum_i \sum_k\left(\Omega_{i}^{k}\right)^{(2)} \vec{c}_i \vec{c_i}
\end{equation}

The surface stress boundary condition at the interface between two fluids 
can be expressed as follows (\cite{landau1959course,brackbill1992continuum}):

\begin{equation}
\mathbf{T}^1 \cdot \vec{n} - \mathbf{T}^2 \cdot \vec{n} = \sigma (\nabla \cdot {\vec{n}}) \vec{n} - \nabla \sigma
\end{equation}

where, $\mathbf{I}$ is the identity tensor, $\sigma$ is the surface 
tension coefficient (with dimensions of force per unit area), $\vec{n}$ is the unit normal to the interface, $\mathbf{T}=-p\mathbf{I} + \rho\nu(\nabla\vec{u} + \nabla \vec{u}^T)$ is the stress tensor of the $k^{th}$ component and $\nabla \cdot \vec{n}$ is the local curvature of the fluid interface.\\
The local stress jump at the interface can be induced by adding an interfacial volume force $\vec{F}(\mathbf{x},t)$ (\cite{liu2012three}):

\begin{equation}\label{force}
\vec{F}(\vec{x},t)= \nabla \sigma \delta_{I} - \delta_{I} [\sigma (\nabla \cdot \vec{n}) \vec{n}],
\end{equation}

In the above, $\delta_{I}=\frac{1}{2} | \nabla \Theta | $ is an index function which explicitly 
localizes the force on the interface and $\Theta=\frac{\rho^1 - \rho^2}{\rho^1 + \rho^2}$ is 
the phase field (\cite{liu2012three}).
The normal to the interface can be approximated by the gradient of the phase field, $\mathbf{n}= \nabla \Theta/|\nabla \Theta|$.

Since the perturbation operator is responsible for generating interfacial tension, the following 
relation must hold:

\begin{equation} \label{SeqF}
\nabla \cdot \bm{\Sigma}= \vec{F}
\end{equation}

By choosing (\cite{leclaire2012numerical}) $\left(\Omega_{i}^{k}\right)^{(2)}= \frac{A_k}{2} |\nabla \Theta|\left[w_i \frac{(\vec{c}_i \cdot \nabla \Theta)^2}{|\nabla \Theta|^2} -B_i \right]$, substituting it into \ref{consconstr} and \ref{SeqF} and by imposing that the set $B_i$ must satisfy the following isotropy constraints:

\begin{eqnarray}
\sum_i B_i= \frac{1}{3} \; \sum_i B_i \vec{c}_i=0 \; \sum_i B_i \vec{c}_i \vec{c}_i= \frac{1}{3} \mathbf{I}
\end{eqnarray}

we obtain an equation for the surface tension of the model:
\begin{equation}\label{sigmaA}
\sigma=\frac{2}{9}(A_1+A_2)\frac{1}{\omega_{eff}}=\frac{4}{9}A\frac{1}{\omega_{eff}}
\end{equation}
The above relation shows a direct link between the surface tension and the parameter $A=A_1+A_2$ ($A_1=A_2$).
\textcolor{black}{In actual practice, after choosing the viscosity of the two components and the surface tension of the model, at each time step, one locally computes the $A=A_1+A_2$  coefficient by using the formula reported in eq. (\ref{sigmaA})}.

\textcolor{black}{It is worth noting that, in this work, a fourth order isotropic discrete gradient operator on a 27 points stencil is employed (for details please refer to \cite{leclaire2017generalized}).}

As pointed out above, the perturbation operator generates an interfacial tension in compliance with the capillary-stress tensor of the Navier-Stokes equations for a multicomponent fluid system.

Nonetheless, the perturbation operator alone does not guarantee the immiscibility of different fluid components.
For this reason, a further step is needed (i.e. the recoloring step) to minimize the mutual diffusion between components.

Following the work of Latva-Kokko and
Rothman (\cite{latva2005diffusion}), the recoloring operator for the two sets of distributions takes the following form:

\begin{eqnarray}
f_i^1=\frac{\rho^1}{\rho} f_i^* + \beta \frac{\rho^1\rho^2}{\rho^2} \cos{\phi_i} f_i^{eq,0} \\
f_i^2=\frac{\rho^2}{\rho} f_i^* - \beta \frac{\rho^1\rho^2}{\rho^2} \cos{\phi_i} f_i^{eq,0}
\end{eqnarray}

where, $f_i^*=\sum_k f_i^{k,*}$ denotes the set of post-perturbation distributions, $\rho=\rho^1 + \rho^2$,  $\cos{\phi_i}$ is the angle between the phase field gradient and the $i^{th}$ lattice vector and $f_i^{eq,0}=f_i(\rho,\vec{u}=0)^{eq}=\sum_k f_i^k(\rho,\vec{u}=0)^{eq}$ is the total zero-velocity equilibrium distribution function (\cite{leclaire2012numerical}).

Note that the coefficient $\beta$ in the above expressions is a free parameter which can be used 
to tune the interface width, thus playing the role of an inverse diffusion length scale (\cite{latva2005diffusion}).
\textcolor{black}{We wish to point out that the present model is employed to simulate droplet-based microfluidic applications which are often characterized by very small $We$, $Re$ and $Ca$ numbers (i.e. much smaller than one).
Hence, by considering  typical flow speeds of $10^{-3}-10^{-2}lu/step$ the $u^3$ error contribution is of the order $O(10^{-12}-10^{-9})$, which reflects in a very negligible compressibility  errors.
}
\textcolor{black}{It is important to note that, following the work of  \cite{leclaire2012numerical},in this work we perform the entire collision step (collision+perturbation+recoloring steps) on two separate distibutions, this at variance with the works of  \cite{gunstensen1991lattice} and \cite{latva2005diffusion}, in which the collision and perturbation are written in terms of the blind distribution $f=f^1 + f^2$.}\\
The color gradient LB scheme is further regularized by filtering out the high-order non-hydrodynamic (ghost) modes, emerging after the streaming step (see (\cite{montessori2015lattice,zhang2006efficient,latt2006lattice,coreixas2017recursive,mattila2017high,hegele2018high,montessori2016effects}) for further details).

Indeed, it was noted that sizeable non-isotropic effects arise in the model (\cite{montessori2018regularized}), whenever the LB scheme is under-relaxed ($\tau \geq 1$).
As a consequence, we exploit the regularization procedure in order to recover the loss of isotropy by
suppressing the non-hydrodynamic modes (\cite{BENZI1992145,montessori2018regularized,montessori2018elucidating}).
\textcolor{black}{For the sake of clarity, here we report a pseudocode of the regularization procedure employed in our simulations:\\
\noindent\fbox{%
\begin{varwidth}{\dimexpr\linewidth-2\fboxsep-2\fboxrule\relax}
\textbf{Regularization step}:
\begin{algorithmic}
\For{$l\leq b \land \forall (i,j,k) \in D$} 
\State $f^{neq,m}_l(i,j,k) = f^{pc,m}_{l}(i,j,k) - f^{eq,m}_l(i,j,k)$  
\EndFor
\For{$l\leq b \land \forall (i,j,k) \in D$} 
\State $p^m_{\alpha\beta}(i,j,k)=p^m_{\alpha\beta}(i,j,k) +  (c_{l\alpha}c_{l\beta} - c_s^2\delta_{\alpha\beta})f^{neq,m}_l(i,j,k)$  
\EndFor
\For{$l\leq b \land \forall (i,j,k) \in D$} 
\State $f^{reg,m}_{l}(i,j,k)= f^{eq,m}_l(i,j,k) +  \sum_l \frac{w_l}{2 c_s^4}(c_{l\alpha}c_{l\beta}-c_s^2)p^m_{\alpha\beta}(i,j,k) $  
\EndFor
\end{algorithmic}
\end{varwidth}
}\\
where $f_l^{pc,m}(i,j,k)$ is the set of post-collision distribution functions of the $m^{th}$ component, $f_l^{neq,m}(i,j,k)$ is the non-equilibrium part of  $f_l^{pc,m}(i,j,k)$, $p^m_{\alpha\beta}$ are the components of the non-equilibrium part of the momentum flux tensor,$D$ stands for the fluid domain ($(i,j,k) \in D$ is a lattice node in the fluid domain) and $f_l^{reg}(i,j,k)$ is the regularized set of post-collision distribution functions}.\\
In the next subsection, we show how to include the effect of repulsive near contact interactions, directly within the LB framework, by augmenting the regularized color-gradient model with a forcing term aimed at coarse-graining the near-contact forces at the fluid surface. 

\subsection{Augmented LB model for repulsive near contact interactions}\label{NCI}

\begin{figure}
\begin{center}
\includegraphics[scale=0.7]{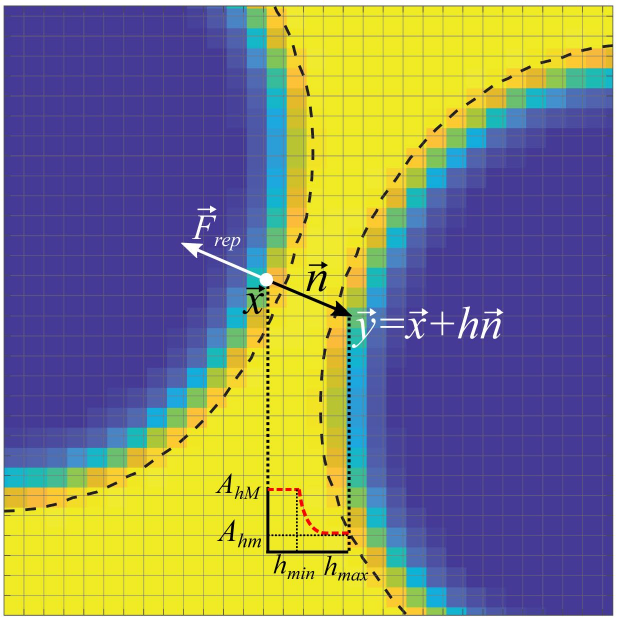}
\caption{\label{sketchrep} Graphical representation of the near interaction between two impacting droplets. }
\end{center}
\end{figure}

The stress-jump condition across a fluid interface is augmented with a repulsive term aimed at 
providing a mesoscale representation of all the repulsive near-contact forces (i.e., Van der Waals, electrostatic, steric and hydration repulsion) acting on much smaller scales ($\sim  O(1 \; nm)$)  than those resolved on the lattice (typically well above hundreds of nanometers).
It takes the following form:
\begin{equation}
\mathbf{T}^1\cdot \vec{n} - \mathbf{T}^2 \cdot \vec{n}=-\nabla(\sigma \mathbf{I} - \sigma (\vec{n}\otimes \vec{n})) - \pi \vec{n}
\end{equation}
where $\pi[h(\vec{x})]$ is responsible for the repulsion between neighboring fluid interfaces, 
$h(\vec{x})$ being the distance along the normal $\vec{n}$, between locations 
$\vec{x}$ and $\vec{y}=\vec{x}+ h \vec{n}$ at the two interfaces, respectively (see Fig. \ref{sketchrep}).\\

The above expression can be rewritten in the following form (\cite{li2016macroscopic}):
\begin{equation}
\mathbf{T}^1\cdot \vec{n} - \mathbf{T}^2 \cdot \vec{n}=\sigma (\nabla \cdot \vec{n})\vec{n} - \nabla_s\sigma - \pi \vec{n}
\end{equation}
\textcolor{black}{in which $\nabla_s$ is used to identify the gradient tangent to the interface.}

By neglecting any variation of the surface tension along the interface, we can approximate  $\mathbf{T}=-p\mathbf{I}$ (\cite{brackbill1992continuum}) and the above equation takes the following form:

\begin{equation}
(-p_1\mathbf{I}) \cdot \vec{n}- (-p_2\mathbf{I})\cdot \vec{n}=\sigma (\nabla \cdot \vec{n})\vec{n}  - \pi \vec{n}
\end{equation}

By projecting the equation along the normal to the surface, we obtain the  \textit{extended} Young-Laplace equation (\cite{chan2011film, williams1982nonlinear}):

\begin{equation}
(p_2 - p_1)=\sigma (\nabla \cdot \vec{n}) - \pi
\end{equation}

The additional term can be readily included within the LB framework, by adding a 
forcing term acting only on the fluid interfaces in near contact, namely:

\begin{equation}
\vec{F}_{rep}= \nabla \pi := - A_{h}[h(\vec{x})]\vec{n} \delta_I
\end{equation}
 
In the above, $A_h[ h(\vec{x})]$ the parameter controlling the strength (force per unit volume)
of the near contact interactions  (please refer to the sketch in figure \ref{sketchrep}).\\
In this work, $A_h$ is set to a constant value ($A_{hM}$) if $h<h_{min}$ and then decreases as $\sim h^{-3}$, as shown in figure \ref{sketchrep}, although other choices are certainly possible. \\
The near contact force has been defined solely as a function of the distance between two fluid interfaces.  Nonetheless, it could be easily extended to take into account local variations due to the effect of spontaneous migrations of the surfactant along the fluid interface (\cite{gupta2017general}).\\
The addition of the repulsive force, naturally leads to the following (extended) 
conservation law for the momentum equation:
\begin{eqnarray} \label{NSEmod}
\frac{\partial \rho \vec{u}}{\partial t} + \nabla \cdot {\rho \vec{u}\vec{u}}=-\nabla p + \nabla \cdot [\rho \nu (\nabla \vec{u} + \nabla \vec{u}^T)] + \nabla \cdot (\bm{\Sigma}  +   \pi \mathbf{I})
\end{eqnarray}
This is the Navier-Stokes equation for a multicomponent system augmented with a surface-localized repulsive term, expressed through the potential function $ \nabla \pi$.

\textcolor{black}{There have been other attempts to model interface interactions in amphiphilic fluids in the literature (see for example \cite{nekovee2000lattice,love2001three,Chen20002043}). These  models aim at directly simulating the effect of a surfactant by evolving two sets of distribution fucntions, allowing to  take into account the presence of an amphiphilic fluid at the interface between two liquid phases.
Indeed, the propagation of the amphiphilic molecules is described a set of LB equations, one for the distribution function  and one for the relaxation of the average dipole vector  to its local equilibrium orientation.
\cite{halliday2007pre} and \cite{spencer2010lattice}  extended the color gradient model  to any number of components. This method makes use of a set of distributions for each immiscible fluid species. 
More recently,\cite{wohrwag2018ternary} proposed a thermodynamically consistent free energy model for fluid flows comprised of one gas and two liquid components using the entropic lattice Boltzmann scheme. \\
The standpoint of our work is quite different in  that, the repulsive action of a surfactant, arising when two interfaces come in close contact, is taken into account by introducing a repulsive forcing term localized at the interface of the two fluids. Importantly, this just requires two distributions functions regardless of the number of droplets.
}

To conclude, the extended approach still holds to a
continuum description of the interface dynamics, being the governing equations 
modified only by the presence of a distributed body force, which can heuristically be interpreted as a coarse-grained version of the short-range molecular forces acting at the nanometers and sub-nanometer scales.

\section{Numerical Results}\label{results}

In this section we test the extended LB model on three applications namely, head-on and off-axis collision between two bouncing droplets, pressure driven flow of a dense emulsion in a channel and the formation of \textit{soft flowing crystals} in a microfluidic flow focuser.

\subsection{Droplets coalescence}

\begin{figure}
\begin{center}
\includegraphics[scale=0.5]{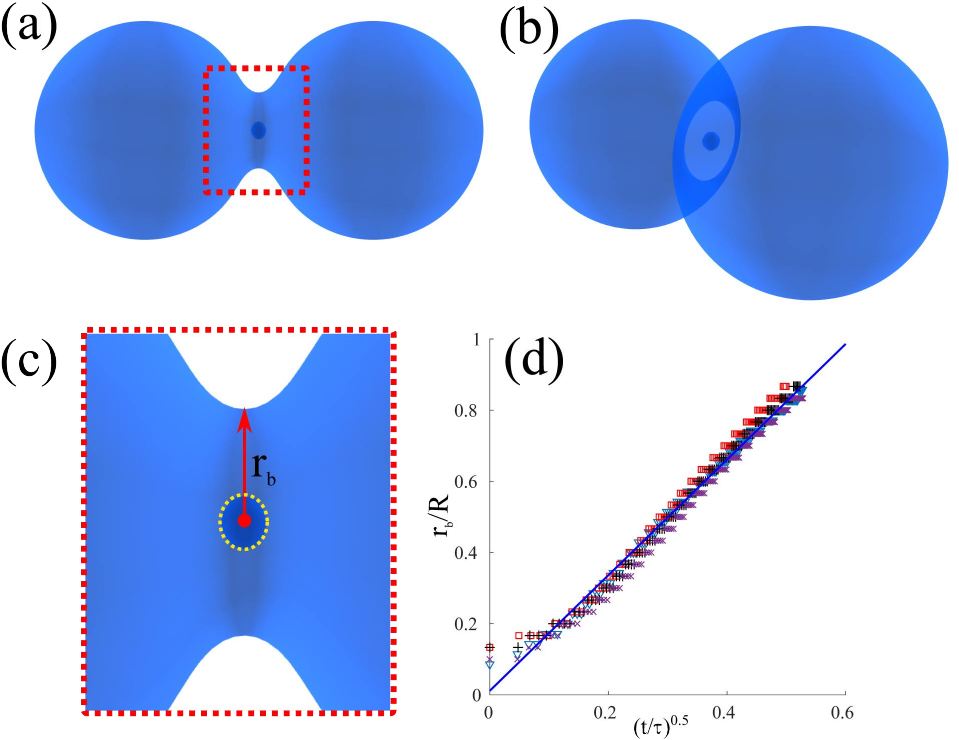}
\caption{\label{egger} \textcolor{black}{(a-c) Bubble formation during the droplets coalescence process due to the fact that only a fraction of the fluid caught
in the narrow gap between the two spheres is able to escape. The rest accumulates in a 'bubble',that forms at the meniscus (\cite{eggers1999coalescence}). (d) normalized bridge radius as a function of the non-dimensional time $t/\tau$, where $\tau=\sqrt{R^3/\sigma}$, for different values of the kinematic viscosity (symbols). In agreement with (\cite{eggers1999coalescence}), the liquid bridge radius $r_b$ is shown to follow a scaling law $r_b\propto t^{1/2}$ with a dimensionless prefactor of $1.6$, in close agreement with the value $1.62$, reported in (\cite{duchemin2003inviscid}).}  }
\end{center}
\end{figure}

\textcolor{black}{Here we first test the ability of the color-gradient LB model to capture the physics of the coalescence process between two equally sized droplets.
The simulation setup consists of a three-dimensional fully periodic box ($160 \times 100 \times 121$) in which two liquid droplets of radius $R=30$ , surrounded by a dispersed fluid of the same density, are placed at close distance. 
The surface tension of the mixture was fixed at a constant value, $\sigma=0.01$ while the surface tension of the two fluids was varied from $\nu=0.05$ to $\nu=0.15$. The viscosity ratio between the droplets and the surrounding phase has been set to unity, and all the physical quantities are reported in lattice units. As shown in fig.\ref{egger} (panel a-c), a liquid bridge forms between the two droplets. As pointed out in (\cite{eggers1999coalescence}) the coalescence process in the early stage is so fast that  only a fraction of the fluid caught
in the narrow gap between the two spheres is able to escape, while the rest accumulates in a "bubble" which forms at the meniscus. The simulations predict the formation of such a bubble, which remains trapped between the two coalescing  droplets.
To be more quantitative, we measured the normalized radius of the liquid bridge, $r_b/R$, as a function of the square root of the non-dimensional time $t/\tau$ (panel d), being $r_b$ the bridge radius, $t$ the simulation time and $\tau$ a characteristic time scale defined as $\tau=\sqrt{R^3/\sigma}$. As evidenced in the figure, the growth of the liquid bridge follows a scaling law $r_b \propto (t/\tau)^{0.5}$, with the data collapsing on a single master curve $r_b/R=1.6\sqrt{t/\tau}$. It is also worth noting that, the prefactor predicted by the simulations ( $1.6$), is in very close agreement with the value $1.62$ reported in (\cite{duchemin2003inviscid}).
}

\subsection{Bouncing colliding droplets}

In this subsection, we show the capability of the extended LB model to accurately reproduce the correct dynamics of head-on and off-axis collisions between two bouncing droplets (\cite{Chen2006}).
\begin{figure}
\begin{center}
\includegraphics[scale=1.2]{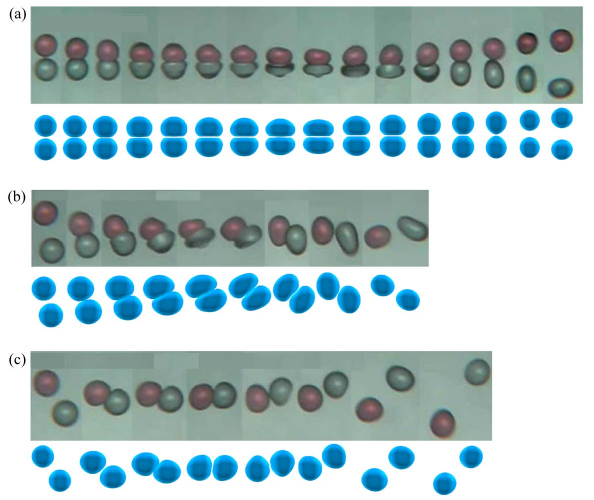}
\caption{\label{collsequence} Collision  sequences for three different impact numbers at different Weber.
(a): $b=0$ and $We=10$. (b): $b=0.33$ and $We=10$. (c): $b=0.85$ and $We=7$. Upper row experiments (\cite{Chen2006}), lower row simulation results.}
\end{center}
\end{figure}
\begin{figure}
\begin{center}
\includegraphics[scale=0.9]{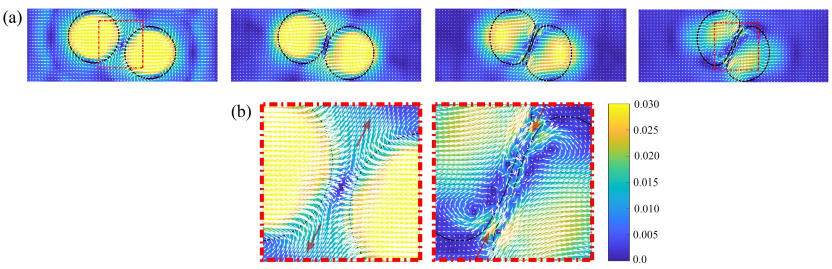}
\caption{\label{thinflow} Sequence of the flow field during the impact between the two droplets (mid-plane slice). As shown in panel (b), during the first stage of the collision the dispersed phase flows outwards, allowing the two droplets to approach. Afterwards, when the droplets come in close contact, the fluid within the thin film begins to recirculate, thereby preventing film rupture hence
the coalescence between the droplets. \textcolor{black}{ A remark is in order: in real interacting systems the thin film between two interfaces in near contact develops on characteristic lenghscales of the order of the nanometers, far below the spatial scales accessible to our simulations. Indeed, in panel (b), the smallest spatial scale is around $20 \mu m$ while the characteristic distance between two non coalescing impacting droplets is of the order of the tenth of nanometers, i.e. three orders of magnitude smaller than the grid size}.}
\end{center}
\end{figure}
The characteristic non-dimensional parameters, governing the collision outcome, are the Weber and the Reynolds numbers, defined as $We = \rho U^2_{rel} D/\sigma$ and
$Re = U_{rel}D/\nu$, respectively.
In the above, $U_{rel}$ the relative impact velocity, $D$ the droplet diameter, $\sigma$ the surface tension coefficient and $\nu$ the kinematic viscosity, as well as the impact number $b=\chi/D$, namely, the 
distance between the collision trajectories in units of the droplet diameter.
In Table \ref{tab1}, we report the main simulation parameters (expressed in lattice units).\\
\textcolor{black}{Also in this case, the viscosity ratio between the droplets and the surrounding phase has been set to unity}.
Figure \ref{collsequence} shows three collision  sequences for different impact, Weber and Reynolds numbers.
The collision outcomes are compared with those reported in (\cite{Chen2006}).
The experiments were performed with near millimetric droplets of immiscible fluids, with diameters ranging between $700-800 \mu m$ and impact velocities in the range of $1-3 m/s$.  
The other relevant parameters can be found in (\cite{Chen2006}).
By taking a droplet diameter of $700 \mu m$, discretized with $30$ lattice units, we obtain 
a lattice spacing $\Delta x\approx 20\mu m = 1 lu$, which is the minimum interaction 
distance between the simulated droplets.\\
\textcolor{black}{It is worth noting that, the thin film between two interfaces in near contact has characteristic lengthscales of the order of  nanometers, far below the spatial scales accessible to our simulations. Indeed, in our simulations, the smallest spatial scale is around 20 $\mu m$ ($\Delta x$), while the characteristic distance between two non coalescing impacting
 droplets is of the order of one to ten nanometers, i.e. three orders of magnitude smaller than the grid size. Nothwithstanding this gap, by matching the main governing parameters (Weber and Reynolds), the overall impact dynamics is correctly captured by the simulations. This, again, calls into cause the universality of the underlying physical processes, i.e., at the spatial scale at hand, the interaction physics depends upon the dimensionless numbers, measuring the relative strength of the interactions, rather than on the strength of the interactions themselves.}
Panel (a) reports the sequence of an head-on collision between two equally sized droplets. \\
As expected, at these Weber and Reynolds numbers, the two droplets bounce off 
without coalescing, because the coalescence is frustrated by the effect of the 
near-contact repulsive forces.

\begin{table}
\begin{center}
\begin{tabular}{|l|l|l|l|l|l|l|l|l|l|}
\hline
   & $nx \times ny \times nz$    & $D$  & $U_{rel}$ & $\sigma$ & $A_h$ & $\nu$ & $b=\chi/D$ & $We$ & $Re$  \\ \hline
a) & $121 \times 101 \times 121$ & $30$ & $0.06$       & $0.01$    & $0.01$               & $0.0167$              & $0$                       & $10$ & $108$ \\ 
b) & $121 \times 101 \times 121$ & $30$ & $0.06$       & $0.01$    & $0.01$              & $0.0167$              & $0.33$                    & $10$ & $108$ \\ 
c) & $121 \times 101 \times 121$ & $30$ & $0.05$       & $0.01$   & $0.01$               & $0.0167$              & $0.85$                    & $7$  & $90$  \\  \hline
\end{tabular}
\caption{\label{tab1} Droplets collision: simulation parameters (lattice units). From the first column on the left: number of computational nodes, droplets diameter, relative impact velocity, surface tension, strength of the near contact force, kinematic viscosity, impact number, Weber and Reynolds numbers. }
\end{center}
\end{table}
The bouncing collision also occurs by increasing the impact parameter (panel (b) and (c)). Indeed, in both cases, the impact velocity is not sufficient to break the intervening thin 
film and a kissing-like collision is finally observed.\\
We then inspected the evolution of the thin liquid film during the collision process. As reported in fig. \ref{thinflow}, in a first stage, the fluid between the two droplets flows outwards, 
thus allowing the  droplets to approach each other (panel b, left).
Afterwards, as they get closer, the liquid begins to recirculate inwards within the intervening film, stabilizing it and preventing the coalescence between the colliding droplets (panel b, right). \\
This phenomenon resembles the so-called \textit{Marangoni flow} in liquid films, namely a fluid recirculation occurring in liquid thin films in the presence of shear and temperature gradients, which was observed to prevent the coalescence between bodies of liquids (\cite{dell1996suppression}) .\\
It is straightforward to note that, by varying the magnitude of the repulsive force, it is possible 
to promote or inhibit the coalescence of the impacting droplets. \\
From this standpoint, it proves expedient to introduce two further 
non-dimensional parameters, which measure the relative strength of  
inertia and surface tension versus repulsive forces.
The first, ($S_\sigma$) is defined as the ratio between  the repulsive force ($\sim A_{hM}$) which frustrates the coalescence between droplets, and the surface tension ($\sigma$), which promotes it.
The second ($S_\kappa$), is defined as the ratio between the local impact kinetic energy  
and the work done by the repulsive forces to prevent the two interfaces from coalescing.

The two non-dimensional parameters read as follows:
\begin{equation}
S_\sigma=(A_{hM} \cdot \Delta x)/\sigma 
\end{equation}
\begin{equation}
S_\kappa=\rho U ^2/(A_{hM} \cdot \Delta x) 
\end{equation}
\\
\begin{figure}
\begin{center}
\includegraphics[scale=0.75]{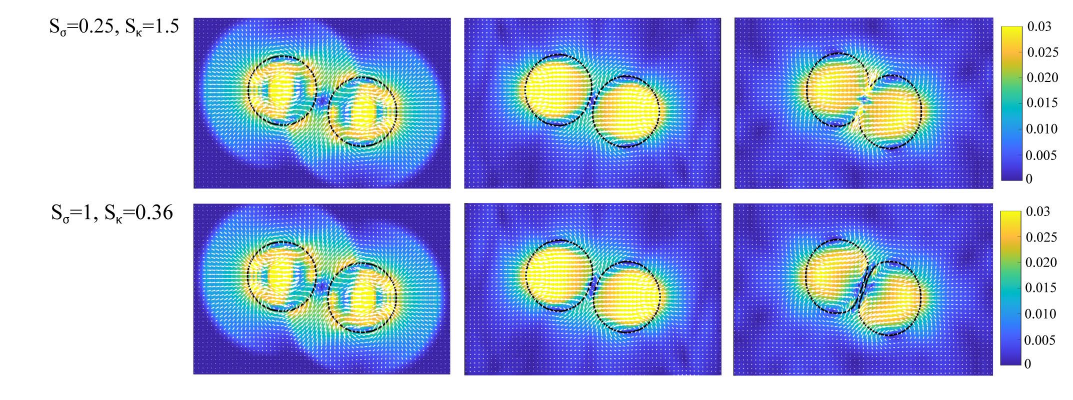}
\caption{\label{comparison} Effect of the magnitude of the near-contact force. The upper panel reports a flow field sequence of two impacting droplets (mid plane slice). The simulation parameters are the same as those reported in Table \ref{tab1}, except for the repulsive parameter, which is four times smaller. 
It is evident that the repulsive force is not strong enough to frustrate the coalescence between the two impacting droplets, as instead occurs in the case reported in the lower panel ($S_{\sigma}=1, S_{\kappa}=0.36$). }
\end{center}
\end{figure}
In the above $\Delta x$, the lattice spacing, is the characteristic length 
scale of near-contact interactions between two droplets on the lattice. 
In our simulations $S_\sigma=1$ and $S_\kappa=0.36$, meaning that the inertial 
and the surface forces (both promoting coalescence) are not strong enough to 
withstand the effect of the local repulsion. \\
By lowering the intensity of the repulsive forces of a factor four  ( $S_\kappa=1.5$ 
and $S_\sigma=0.25$), we finally observe the coalescence between the impacting droplets (see fig. \ref{comparison}). It will be shown below (subsection \ref{softflowing}) that the balance between inertia, surface forces and repulsive actions crucially affects the formation and the 
overall dynamics of crystal-like structures in foams and emulsions.

\subsection{Dense emulsion in a planar channel}

\begin{figure}
\begin{center}
\includegraphics[scale=0.7]{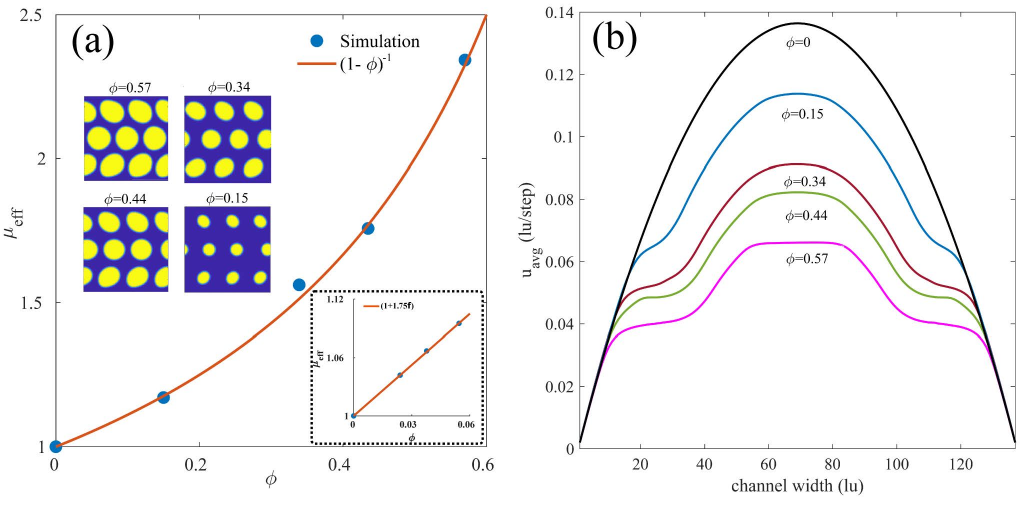}
\caption{\label{murel}Left panel: measured $\mu_{eff}$ (symbols) as a function
of droplets volume fraction $\phi$. The solid line is a fit based on the differential effective medium theory (\cite{bullard2009comparison}). \textcolor{black}{In the inset, we report also the Taylor theory ($\nu_d/\nu_c=1$),(solid line) ($1+1.75\phi$), which holds for small values of the volume fraction,  versus the simulations.}  Right panel: averaged velocity profiles as a function of the packing fraction $\phi$ }
\end{center}
\end{figure}

We now discuss the pressure-driven flow of a dense emulsion within a 
narrow channel, made of a regular arrangement of equal sized droplets 
(component A) dispersed in a continuous matrix (component B).
\begin{figure}
\begin{center}
\includegraphics[scale=0.5]{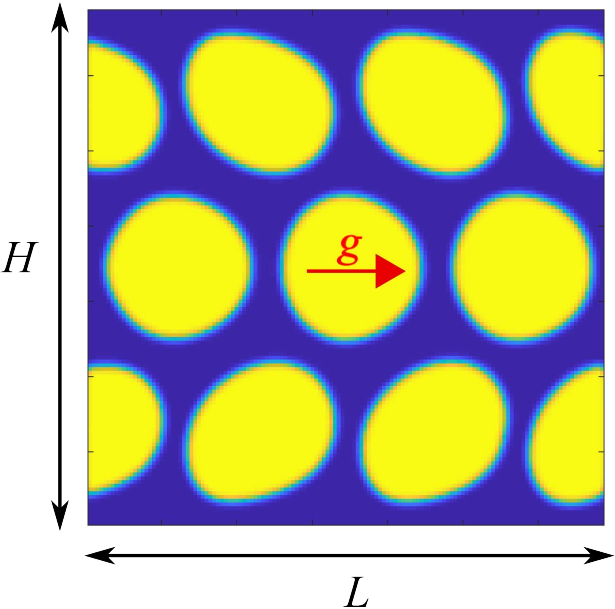}
\caption{\label{sketchemu} Sketch of the pressure driven flow of an emulsion in a narrow channel. $H \times L=137 \times 137 $ while $g$, the applied force employed to mimic the pressure gradient along the channel, is set to $10^{-5}$, to guarantee laminar flow conditions. }
\end{center}
\end{figure}
The simulations are performed on a $137 \times 1 \times 137$ ($H\times W \times L$) nodes domain. \textcolor{black}{The viscosities of both the dispersed and the continuous phase are set to $0.167$}, while the surface tension is set to $\sigma=0.02$ (both in lattice units). Periodic boundary conditions have been applied cross-flow and along the flow direction while, on the upper and lower walls, no-slip conditions have been imposed. The static contact angle between the dispersed phase and the solid walls is set  to $180^\circ$ (pure hydrophobic walls).
A body force ($g=10^{-5}$, in lattice units) is employed to mimic the effect of an applied pressure gradient along the channel. The value of $g$ was chosen so as to keep the Reynolds number 
sufficiently low ($Re \leq 10$) to guarantee laminar flow conditions 
within the channel, as typical of microfluidic channels.\\
Figure \ref{murel} (right panel) reports the averaged velocity profiles at different values of the packing fraction, $\phi=n\cdot V_{drop}/V_{tot}$, being $n$ the number of droplets in the channel, $V_{drop}$ the volume of the (cylindrical) droplet and $V_{tot}$ the volume of the channel.
When $\phi=0$, the usual Poiseuille flow parabolic profile for a pure fluid is recovered, 
as shown in the left panel of fig. \ref{murel}.

As $\phi$ increases, the velocity profiles flatten in the central region of the channel.
In order to quantify the effect of the packing fraction, we measured the effective (or relative) dynamic viscosity, defined as the flow rate ratio $\mu_{eff}=Q(\phi=0)/Q(\phi)$ and compared it against the model \textcolor{black}{proposed by   \cite{taylor1932viscosity}} and  \cite{bullard2009comparison}.
\textcolor{black}{Taylor theory, the effective dynamic viscosity (for $\nu_d/\nu_c=1$) can be expressed as a linear function of the volume fraction (in the limit of small droplets and volume fractions) as follows:
\begin{equation}
\mu_{eff}=1+1.75\phi
\end{equation}}
The effective dynamic viscosity predicted by Bullard, which is based on the differential effective medium theory, read as follows:
\begin{equation}
\mu_{eff}=(1-\phi)^{-[\eta]}
\end{equation}
being $[\eta]$ the  intrinsic viscosity, whose value is restricted between the undeformable and the freely deformable limit (\cite{douglas1995intrinsic}).
As reported in figure \ref{murel} (left panel), the simulation results are in a very close match with respect to the theoretical prediction of Taylor (inset in left panel) and Bullard (with intrinsic viscosity set to $[\eta]=1$). 

\subsection{ Soft flowing crystals in a microfluidic focuser} \label{softflowing}

\begin{figure}
\begin{center}
\includegraphics[scale=0.8]{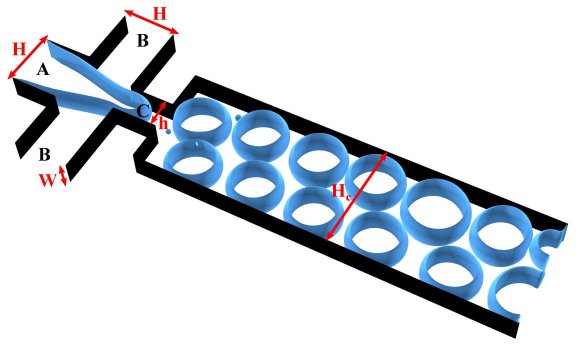}
\caption{\label{sketchff}  Sketch of the flow focuser device. }
\end{center}
\end{figure}

As an application, we simulated the formation of oil/water emulsions 
in a microfluidics flow focusing device, whose sketch is reported in fig. \ref{sketchff}.
The micro-device is made of three channels supplying the dispersed (A) and the 
continuous (B) phase, plus an orifice (C) placed downstream of the three coaxial inlet streams.

The mechanism of droplet formation follows from the periodic pinch-off of the dispersed jet by the continuous stream and the pinch-off mechanism takes place in the small orifice.

\begin{figure}
\begin{center}
\includegraphics[scale=0.7]{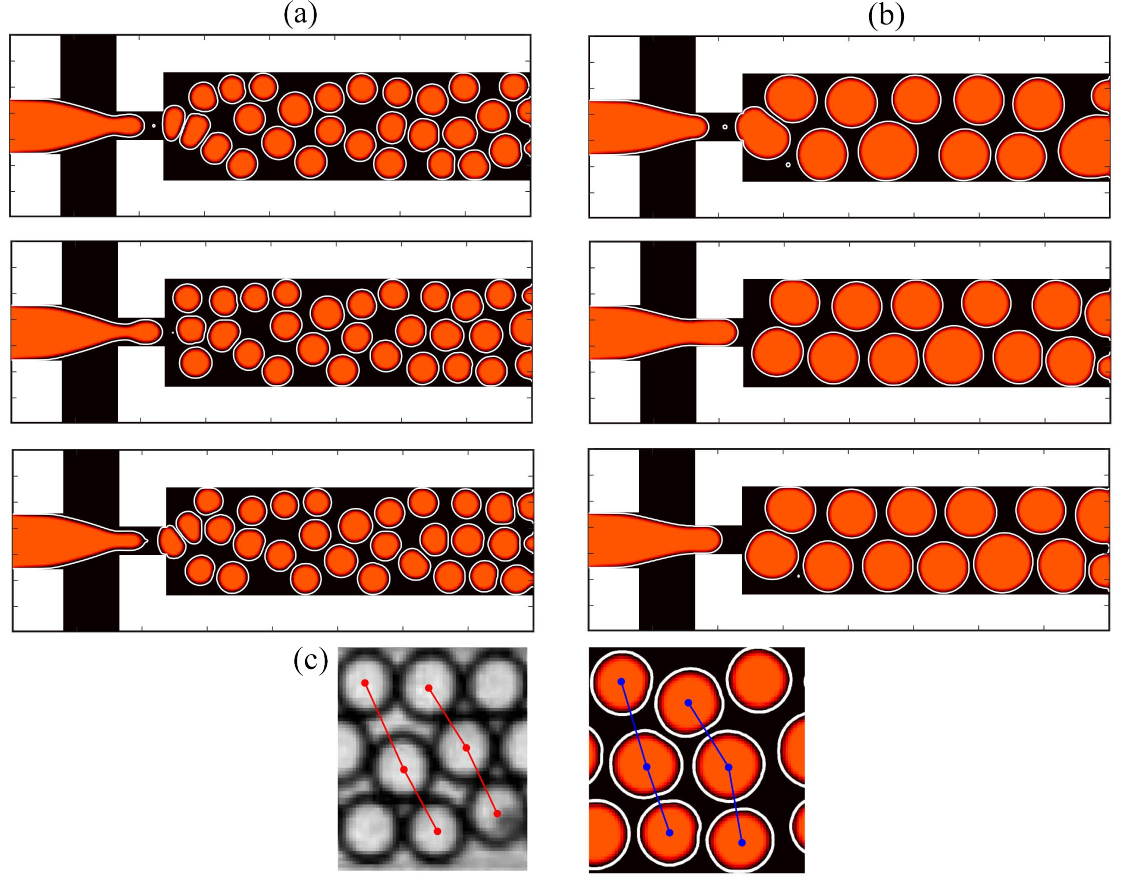}
\caption{\label{flow1} (a -b) Different configurations of monodispersed emulsions at the outlet of the flow-focuser, on an $xy$ midplane, obtained by changing the dispersed/continuous flowrate ratios. (c) Zoom of the three droplets foam structure: visual comparison between experiments (Marmottan et al. (\cite{marmottantprl2009}), left panel) and numerical simulations (right panel). }
\end{center}
\end{figure}

\begin{figure}
\begin{center}
\includegraphics[scale=0.7]{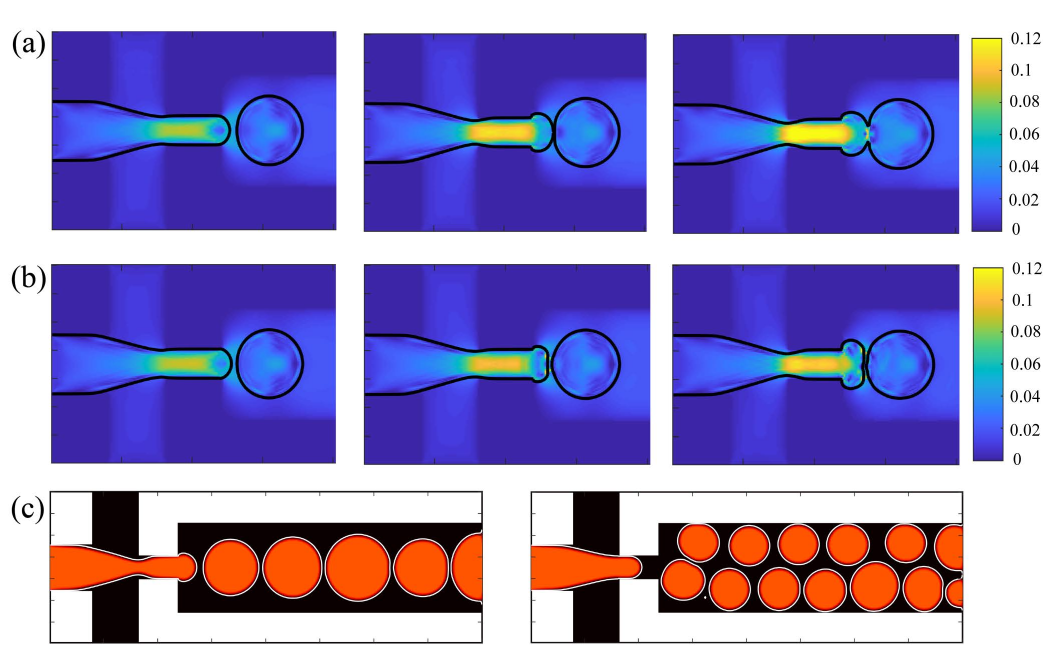}
\caption{\label{ffcoalescence} (a-b) Flow field sequence during the break up stage occurring at the outlet of the striction of the flow focuser. (a) lower (b) higher near contact force. In the first case the strenght of the repulsive force is not sufficient to prevent the coalescence between the droplet and the jet, due to the high speed of the jet at the outlet of the nozzle. Nonetheless, the repulsive force is strong enough to prevent the coalescence between droplets moving in the outlet channel (see panel (c), left figure). In the second case, the repulsive force is strong enough to completely suppress the coalescence between the jet and the newly formed droplets, thus allowing the formation 
of an ordered soft flowing crystal in the outlet channel.}
\end{center}
\end{figure}

Flow focusers are nowadays widely employed for the production of mono-dispersed 
emulsions, due to the precise control over the emulsion monodispersity and the droplet size (\cite{whitesides2006origins,sackmann2014present,cruz2017global}).
The high degree of flow reproducibility is due to the dominance of the viscous forces over inertia 
which smoothes flows and tames hydrodynamic instabilities. 
The pinch-off process is thus metronomic, allowing to  produce
droplets with measured standard deviations in size as little as the $0.1\%$  (\cite{marmottant2009microfluidics,ganan2001perfectly,link2004geometrically}).\\
Here, we show that the mesoscale approach proposed in this paper is able to reproduce different packing configurations in the outlet channel of the flow focuser, which are obtained by varying the dispersed-to-continuous flow rate ratio. Moreover, we also show that, by tuning the magnitude of the near force interaction, different structures of the flowing crystals can be achieved.
The simulation setup is sketched in figure \ref{sketchff}, while the main simulation parameters are reported in Table \ref{tab2}.\\
The prototypical focuser, used in the simulations, consists of three main channels (A and B) of width $H=200 \mu m$, a nozzle (C) $h=100 \mu m$ and an outlet channel of width $H_c=400 \mu m$, while the height of the focuser is $ W=100 \mu m$.
By taking an interfacial tension of an oil-water mixture ($\sim 50 mN/m$), the dynamic viscosity of the water (dispersed phase) $\mu \sim 10^{-3} Pa\cdot s$ and an inlet velocity of the dispersed phase $\sim 0.1 m/s $ we obtain a Weber number $We=0.04$ and a capillary number $Ca=0.0017$, which are typical of flow-focuser devices (\cite{marmottant2009microfluidics}).
Figure \ref{flow1} reports two different flow configurations which can be obtained 
by varying the flow rate ratios between the dispersed and the continuous phase.
In both cases, droplets readily self-assemble in ordered patterns described as \textit{soft flowing crystals} 
(\cite{garstecki2006flowing,marmottant2009microfluidics,dollet_scagliarini_sbragaglia_2015}.
In panel (a) ($\phi=1/2$ ), the reported sequence shows a typical wet foam-like configuration in which the droplets are circular (cylinders in three dimensions) and automatically arrange on three rows, as evidenced in fig.\ref{flow1} panel (c), which also provides a visual comparison with the experimental data reported in (\cite{marmottantprl2009}).
Panel (b) ( $\phi=1/1$) shows a more ordered crystal structure, made 
of larger cylindrical droplets disposed along two staggered rows.\\
We further investigated the effect of the magnitude of the near force 
on the formation of the crystal pattern.
Figure \ref{ffcoalescence} panel (a) shows a time sequence of the flow field during the breakup stage occurring downstream the striction of the focuser, with a magnitude of the near contact force lowered by a factor $5$ with respect to the base case of fig.\ref{flow1} panel (b).  
It is evident that the magnitude of the repulsive force is no longer sufficient to prevent the coalescence between the droplet downstream the striction and the upstream jet, due to the high impact velocity. Thus, as evidenced in panel (c), the droplets in the outlet channel are about 
twice the size with respect to the previous case, and proceed in single-file motion, i.e. aligned 
along the horizontal axes of the focuser.
It is also interesting to note that the action of the repulsive force keeps frustrating the coalescence in the outlet channel, where the typical velocity are much lower than at the exit of the nozzle.
Once again, we can compute the non-dimensional parameter 
$S_\kappa$, downstream of the nozzle, where the coalescence occurs.
By taking a characteristic jet velocity $U_J \sim 0.1$, we estimate $S_\kappa \sim 1.4$ (fig. \ref{ffcoalescence} panel (a)) and  $S_\kappa \sim 0.18$ (fig. \ref{ffcoalescence} panel (b)). 
Thus, $S_\kappa \sim O(1)$ may be interpreted as a threshold above which the repulsive force is no longer able to balance the inertia, so as to frustrate the coalescence between the jet and the droplet. These preliminary simulations clearly highlight the pivotal role of the near contact interactions on the structure of the resulting flowing crystal.\\
\begin{table}
\begin{center}
\begin{tabular}{|l|l|l|l|l|l|l|l|l|l|}
\hline
   & $nx \times ny \times nz$    & $H,W,H_c,h$  & $u^{in}_{d}$ & $u^{in}_{c}$ & $\nu_d$ & $\nu_c$ & $\sigma$ & $A_{hM}$  & $\Gamma$  \\ \hline
(fig.\ref{flow1} (a)) & $400 \times 20 \times 140$ & $40,20,80,20$ & $0.01$       & $0.01$    & $0.0167$               & $0.1167$              & $0.1$                       & $0.1$ & $1/2$ \\ 
(figs.\ref{flow1} (b)(c)-\ref{ffcoalescence}(b)) & $400 \times 20 \times 140$ & $40,20,80,20$ & $0.01$       & $0.005$    & $0.0167$               & $0.1167$              & $0.1$                       & $0.1$ & $1/1$ \\ 
(fig.\ref{ffcoalescence}(a)(c)) & $400 \times 20 \times 140$ & $40,20,80,20$ & $0.01$       & $0.005$    & $0.0167$               & $0.1167$              & $0.1$                       & $0.02$ & $1/1$ \\ \hline
\end{tabular}
\caption{\label{tab2} Flow focuser simulations: main parameters (lattice units). From the first column on the left: number of computational nodes, device characteristics, inlet velocity of the dispersed phase, inlet velocity of the continuous phase, dispersed phase viscosity, continuous phase viscosity, repulsive interaction parameter, flow rate ratio. \textcolor{black}{\textcolor{black}{The viscosity ratio between the droplets and the surrounding phase has been set $\nu_d/\nu_c=1/7$.}} }
\end{center}
\end{table}
From this standpoint, the proposed approach may open up new chances to investigate the complex dynamics of flowing microfluidics crystals, helping in identifying the optimal operational regimes required to precisely control the  production of mono-dispersed emulsions. 

\section{Conclusions}\label{summary}

The dynamic interactions occurring at fluid interfaces at nanometric and 
subnanometric scales, are known to play a crucial role on the macroscopic 
behavior of complex states of densely packed soft flowing matter,
such as colloidal systems, foams and emulsions.
As a result, an in-depth knowledge of these dynamic interactions is pivotal to a deeper understanding of the properties of \textit{soft flowing crystals} (\cite{marmottant2009microfluidics,garstecki2006flowing,montessori2018mesoscale}).
Many theoretical and experimental researchers endeavored to explain the basic physics behind near contact interactions. 
Notwithstanding the surge of experimental and theoretical activity, the numerical description of the \textit{soft flowing matter} is still in its early stage.\\
One reason is that, the concurrent solution of the macroscopic equations, needed to evolve the fluid interface, together with a direct description of the near contact interactions at the 
nano-scales stands out as a prohibitive multiscale problem.

In this paper, we have proposed a coarse-grained approach to include the effect of the near contact interactions within the lattice Boltzmann computational framework, and shown that 
such extended LB model is able to accurately describe a number of relevant physical effects.
That is, i) to capture the evolution of two bouncing impacting droplets for different values of the main governing parameters namely, Weber Reynolds and impact number; ii) to predict the effective viscosity of a dense emulsion flowing in a micro-channel, in agreement with the theoretical model of Bullard (\cite{bullard2009comparison}).
Moreover, the extended LB approach is also able to reproduce different droplets arrangements at the outlet channel of a microfluidic focuser, thus permitting to 
simulate \textit{soft flowing crystals} at the scale of the actual microfluidic device.

To this purpose, two additional non-dimensional parameters ($S_\sigma$ and $S_\kappa$) have been introduced which measure the strength of inertia and surface tension versus the repulsive near-contact interactions.
We found that, $S_{\kappa}=1$ acts as a natural threshold, above which the repulsive near-contact 
forces are no longer able to withstand the impact kinetic energy and prevent the 
coalescence between colliding fluid interfaces.

Even though in this paper we have discussed the specific case 
of a flow-focuser microfluidic device, the method presented in this work  
is expected to apply to a much broader variety of engineering and biomedical problems.

An application where our methods can be specifically useful (and in  future will be developed for) is the design of systems where droplets would undergo an internal transition from viscous solutions to elastic materials. This in-droplet gelation "freezes" the shape of microparticles as it is at the outlet channel. The resulting microparticle systems (also produced in microfluidic devices) are employed e.g. for the encapsulation of living cells in hydrogels (biological reactors or sensors for toxicological screening) or of pharmaceutically active compounds in polymer matrices (controlled drug release). 
By applying our methods to such materials, it will be possible to rationally design complex, micro particle-based flowing crystals, where morphology/aspect ratio and any ensuing properties are precisely and topologically controlled. This would allow e.g. a) a permanent and tailored modulation of optical properties orthogonally to the channel direction, producing soft waveguides, or b) a very fine control of the conditions of dynamic arrest (macroscopic 'gelation') of the emulsion or foam obtained through the microfluidic device, which can be particularly useful in applications of 3D printing. 

\textcolor{black}{We would like to stress that the possibility of employing a mesoscale instead of a full scale description crucially relies upon the universality of the underlying physics or, in other words, its dependence on suitable dimensionless parameters measuring the relative strength of the interactions, rather than on the microscopic details  of the interactions themselves.
The aim of this work was precisely to present a coarse-grained approach encompassing the basic physical features of near contact interactions. In this regard, the proposed model represents an upscale of the interactions occurring at the interface level.
Whilst being a dramatic simplification of the underlying physics at the molecular level, the results obtained in this paper suggest that, at least at the spatial scale at hand, a coarse grained description is appropriate to describe the  mesoscale evolution of an interacting multidroplet system.
}
To conclude, it is hoped that the numerical approach presented here may open the way 
to an experimental-scale modeling of \textit{soft flowing crystals}, promising new 
chances to decode the complexity which characterize this fascinating state of flowing matter.

\section*{Acknowledgement}

The research leading to these results has received
funding from the European Research Council under the European
Union's Horizon 2020 Framework Programme (No. FP/2014-
2020)/ERC Grant Agreement No. 739964 (COPMAT).



\end{document}